\documentclass[aps,pra,twocolumn,showpacs,amsmath,amssymb]{revtex4}

\usepackage{amsmath}
\usepackage{amssymb}
\usepackage{graphicx}
\usepackage{epsf}
\usepackage{flafter}%
\usepackage{bm}

\begin{document}

\title{ Many-body dynamics of Rydberg excitation using the $\Omega$-expansion}

\author{J. Stanojevic}
\author{R. C\^ot\'e}
\affiliation{Department of Physics, University of Connecticut,
Storrs, CT 06269}
\date{\today}

\begin{abstract}
We investigate the excitation dynamics of Rydberg atoms in ultracold atomic samples by expanding the excitation probability and the correlation function between excited atoms in powers of the  isolated atom Rabi frequency $\Omega$. 
In the Heisenberg picture, we give recurrence relations to calculate any order of the expansions, which ere expected to be well-behaved for arbitrarily strong interactions. For homogeneous large samples, we give the explicit form of the expansions, up to $\Omega^4$, averaged over all possible random spatial distributions of atoms, for the most important cases of excitation pulses and interactions. 
\end{abstract}


\maketitle

\section{Introduction}
In recent years, the interactions between ultracold Rydberg atoms have been extensively studied for their possible significance in molecular and ultracold  plasma physic, as well as quantum computing. It has been proposed to use strong interactions between Rydberg atoms to entangle neutral atoms and achieve fast quantum gates \cite{jaksch00,Grangier02}. One effect of these interactions is the excitation blockade, where one excited atom prevents the excitation of nearby atoms. This effect could be utilized to realize scalable quantum gates \cite{lukin01}. In macroscopic samples, the blockade mechanism manifests itself as a suppression of Rydberg excitation. A local blockade of Rydberg state excitation in a mesoscopic sample due to strong van der Waals (vdW) interactions has been observed \cite{tong04} using a pulse-amplified single-mode laser. We have proposed a mean-field type model to explain these experimental results \cite{tong04}. In the model, a distribution of mean-field shifts was constructed for which a distribution of excitation probabilities was calculated. The agreement between the theoretical model and experimental measurements was good. Suppression of Rydberg excitation has also been measured in  two-step excitation processes using cw excitations \cite{singer04,Liebisch,voght06}. The atom counting statistics of Rydberg excitation is significantly modified by the interactions and its sub-Poissonian character has been observed \cite{Liebisch,Ates06}.

Strong Rydberg-Rydberg interactions can impose correlation between atoms within a range of a few $\mu{\rm m}$  so that  many-body treatments are, in principle, necessary. Various treatments \cite{tong04,Hernandez,Pohl07} have been proposed to describe these systems. For example, rate equations were used  for certain  two-step excitation schemes in \cite{Ates07,Pohl07}. However, this approach is not applicable for the systems we primarily consider because spontaneous decay and related decoherence effects are negligible in our case. A different approach is to numerically calculate the many-body wave function \cite{Hernandez,Stanojevic07}. For strong correlations between nearby atoms, the concept of pseudo(super)atoms can be introduced \cite{Hernandez} to reduce the number of many-body states needed for the numerical simulations.  We can numerically evaluate the many-body wave function of mesoscopic systems of $\sim$10 $\mu$m  diameter  \cite{Stanojevic07}. Large systems are difficult to describe in this method, so that the interactions with faraway atoms is modeled by mean-field shifts while the correlations  between nearby superatoms are fully accounted for \cite{Hernandez}. 

In this paper, we show how to calculate the $\Omega$-expansions of the excitation probability and correlation function in the Heisenberg picture. The explicit forms of the lowest orders of these expansions are derived and for large systems their ensemble averaged counterparts are given.

\section{$\Omega$-Expansion in the Heisenberg picture} 
We consider a system of $N$ ultracold two-level atoms for which the upper level is a Rydberg state. 
We assume that the thermal motion of the atoms is greatly reduced  and can be completely ignored, leading to the so called ``frozen'' gas approximation. The interactions between Rydberg atoms are presumably strong so that many-body effects may occur.

We start with the many-body Hamiltonian   of interacting two-level atoms  (with $\hbar=1$)
\begin{eqnarray}\label{Hamiltonian4}
H &=& \Delta \!\sum\limits_{i = 1}^N {\hat \sigma _{ee}^i } \!+\! \frac{\Omega }{2}  \sum\limits_{i = 1}^N  \left( {w(t) \hat \sigma _{eg}^i  +w^{*}(t) \hat \sigma _{ge}^i } \right)\nonumber\\
 &&{}+ \sum\limits_{i = 1,j > i}^N {\kappa_{ij}} \hat \sigma _{ee}^i \hat \sigma _{ee}^j ,
\end{eqnarray}
where $\Delta$ is the frequency detuning from resonance and $\kappa_{ij}$ are the interaction strengths between Rydberg atoms. The second term in the Hamiltonian is the dipole operator representing the interaction with the optical field. The function $w(t)$ is the time evolution (envelope) of the laser pulse.  The  $\sigma$-operators  are defined as 
$\hat \sigma _{\alpha\beta}^i = \left| {\alpha_i } \right\rangle \left\langle {\beta_i } \right|\otimes \hat I^{i(N-1)}$, where $\alpha,\beta$ refer either to the ground state $g$ or the excited state $e$ and  $I^{i(N-1)}$ is the identity operator in the subspace which is an orthogonal complement to $\left|g_i\right\rangle \left\langle g_i \right|\oplus\left|e_i\right\rangle \left\langle e_i \right|$.

Solving this Hamiltonian is trivial in two limits, when all $\kappa_{ij}\!\rightarrow\!\infty$ and when all $\kappa_{ij}\!\rightarrow\!0$.  For $\kappa_{ij}\!\rightarrow\!0$ and real $w(t)$, we get the isolated atom excitation probability given by the Rabi formula $P_{\rm exc}(t)=\sin^2(\Omega W(t)/2)$, where 
$\Omega W(t)=\Omega \int_{t_0}^t w(t')dt'$ is the pulse area and $t_0$ is the initial time of laser excitation.  
For $\kappa_{ij}\!\rightarrow\!\infty$, an arbitrary big ensemble of atoms becomes fully blockaded. This means that there cannot be more than one excited atom in the sample. 
Such systems are effectively  two-level systems, with the collective ground state $\left| G \right\rangle=\left| g_1g_2\ldots g_N \right\rangle$ and the collective excited state $\left| E \right\rangle=1/\sqrt{N}\sum_{i}\left| g_1\ldots e_i\ldots g_N \right\rangle$. Consequently, the equivalent Hamiltonian $H'$ in the limit $\kappa_{ij}\!\rightarrow\!\infty$ is
\begin{equation}\label{strong_inter_Hamiltonian}
\hspace{-0.mm}H'\!= \Delta\hat\sigma_{EE} + \frac{\sqrt{N}\Omega }{2} \left( w(t)\hat\sigma_{EG}\!+\! w^{*}(t) \hat\sigma_{GE}\right).
\end{equation}
This is formally the Hamiltonian for isolated atoms, only  the Rabi frequency scales as $\sqrt{N}$.
If the detuning $\Delta$ is zero and $w(t)$ real, the solution is \cite{wolker}
\begin{equation}\label{strong_inter_solut}
P_{\mathrm {exc}}(t) =
	 \frac{1}{N}{\mathrm{sin}}^{2} \left( \sqrt{N} \Omega W(t) /2 \right) \, .
\end{equation}

We can solve the equations of motion by expanding the $\sigma$-operators in powers of $\Omega$. It follows from Eq. (\ref{strong_inter_solut}) that the expansion is well defined and convergent in the limit  $\kappa_{ij}\!\rightarrow\!\infty$. Clearly, since it is also fine for isolated atoms (the limit $\kappa_{ij}\!\rightarrow\!0$), we expect that it is well defined and convergent for arbitrarily strong interactions.

It is remarkable that, in principle, all terms in the expansion can be calculated exactly. In this sense, the Hamiltonian (\ref{Hamiltonian4}) is exactly solvable. The usual problem with a many-body Hamiltonian is that the nonlinear part containing interactions is very difficult to solve.
Here the nonlinear part $H_0 = \Delta \Sigma_{i = 1}^N {\hat \sigma _{ee}^i }  + \Sigma_{i = 1,j > i}^N {\kappa_{ij}} \hat \sigma_{ee}^i \hat \sigma _{ee}^j $ is exactly solvable.
This Hamiltonian cannot change the number of excited atoms and thus any pure collective 
state with a fixed number of excited atoms is an eigenstate of $H_0$. This is essentially the reason allowing the calculation of any term in the $\Omega$-expansion.

\subsection{$\Omega$-Expansion of the excitation probability} 
The evolution of the $\sigma$-operators in the Heisenberg picture, governed by the Hamiltonian (\ref{Hamiltonian4}), is given by the following equations  
\begin{align}
\frac{d\hat \sigma _{ee}^i}
{d\tau } =& i\frac{\Omega }{2} \left[ {g(t)\hat \sigma _{eg}^i - g^{*}(t)\hat \sigma _{ge}^i } \right] ,\label{dif_eq_sigm_ee}\\
\hspace{-1mm}\frac{{d\hat \sigma _{eg}^i }}{{d\tau }} 
=& i\Delta \hat \sigma _{eg}^i  + i\frac{\Omega }
{2}g^{*} (t)\left[ {2\hat \sigma _{ee}^i \!-\! 1} \right] 
\!+\! i\sum\limits_{j \ne i}\!\kappa_{ij}  \hat \sigma _{eg}^i \hat \sigma _{ee}^j \, \label{dif_eq_sigm_eg}.
\end{align}
These equations can be simplified by removing the $\Delta$ term using new scaled (dimensionless) variables
\begin{eqnarray*}
\tau = t/T , &{\phantom{a}}\omega =\Omega T, &{\phantom{a}}\hat \sigma _{eg}^j e^{-i\Delta t}\to \hat \sigma _{eg}^j,\\
\delta  = \Delta T, &{\phantom{a}}k_{ij}=\kappa_{ij}T,&{\phantom{a}}f(\tau) = g(\tau T)e^{i\delta \tau },
\end{eqnarray*}
where $T$  is the pulse duration. For Gaussian pulses, we use $T=T_{FWHM}/\sqrt{2\ln2}$, where $T_{FWHM}$ is the interval  between the instants for which the laser power is a half of the maximum value.
 The equations for the new $\sigma$-operators can then be rewritten as 
\begin{align}
\frac{d \hat \sigma _{ee}^{i} }{d \tau} =& \frac{i \omega}{2} \left[ {f(\tau )\hat \sigma _{eg}^{ i } - f^{*}(\tau )\hat \sigma _{ge}^{ i }  } \right] ,\label{ee-DJ-second}\\
\frac{d \hat \sigma _{eg}^{i} }{d \tau} =& \frac{i\omega }
{2}f^{*}(t)\left[ {2\hat \sigma _{ee}^i  - 1} \right] 
 + i\!\sum\limits_{j \ne i} {k_{ij} } \hat \sigma _{eg}^i \hat \sigma _{ee}^j \,. \label{eg-DJ-second}
\end{align}

We solve the equations of motion by expanding the $\sigma$-operators in power of $\omega$. 
\begin{eqnarray}
\hat \sigma _{ee}^i & =& \hat \sigma _{ee}^{i(0)}  + \omega \hat \sigma _{ee}^{i(1)}  + \omega ^2 \hat \sigma _{ee}^{i(2)}  + \ldots ,\label{ee-expansion}\\
\hat \sigma _{eg}^i & =& \hat \sigma _{eg}^{i(0)}  + \omega \hat \sigma _{eg}^{i(1)}  + \omega ^2 \hat \sigma _{eg}^{i(2)}  + \ldots .\label{eg-expansion}
\end{eqnarray}

The $\omega$-expansion of $\left\langle {\hat \sigma _{ee}^{i}(\tau )} \right\rangle $ $\left(\left\langle {\hat \sigma _{eg}^{i}(\tau )} \right\rangle\right)$ contains only even (odd) power of $\omega$ if all the atoms are initially in their ground state. This can be shown as follows. Instead of $|g\rangle$ and $|e\rangle$ we can use $|g'\rangle$ and $|e'\rangle$ defined as
\begin{equation}\label{phasetransform}
 \left| g' \right\rangle =e^{i\varphi_g}\left| g \right\rangle \, ,{\phantom{aaaa}} \left| e' \right\rangle = e^{i\varphi_e}\left| e \right\rangle \, ,
\end{equation}
where the constant phase factors satisfy $\varphi_e - \varphi_g =\pm\pi$. We can then define new operators $\hat \sigma _{e'e'}$ and $\hat \sigma _{e'g'} $. It follows from the definitions of these operators that for any $\tau$ and $\omega$ the following relations have to be satisfied 
\begin{equation}\label{sol_sym1}
\hat \sigma_{e'e'}(\tau,\omega)=\hat \sigma_{ee}(\tau,\omega), \quad
\hat \sigma_{e'g'}(\tau,\omega)=-\hat \sigma_{eg}(\tau,\omega),
\end{equation}
which gives
\begin{equation}\label{sol_sym2}
\begin{split}
&\langle\hat \sigma_{e'e'}(\tau,\omega)\rangle=\langle\hat \sigma_{ee}(\tau,\omega)\rangle, \\
&\langle\hat \sigma_{e'g'}(\tau,\omega)\rangle=-\langle\hat \sigma_{eg}(\tau,\omega)\rangle.
\end{split}
\end{equation}
Similarly to Eqs. (\ref{ee-DJ-second}) and (\ref{eg-DJ-second}), we can write evolution equations for the new operators $\hat \sigma _{e'e'} $ and $\hat \sigma _{e'g'} $. We can also write Eqs. (\ref{ee-DJ-second}) and (\ref{eg-DJ-second}) for the opposite sign of omega $\omega\rightarrow -\omega$. However, after the substitution $\omega'=-\omega$,  the time evolution of the expectation values $\langle\hat \sigma_{ee}(t,\omega)\rangle$ and $\langle\hat \sigma_{eg}(t,\omega)\rangle$,
and  the time evolution of $\langle\hat \sigma_{e'e'}(t,\omega')\rangle$ and $\langle\hat \sigma_{e'g'}(t,\omega')\rangle$  are given by the same differential equations. Since the initial conditions are the same, the solutions have to be the same as well 
\begin{equation}\label{sol_sym3}
\begin{split}
\langle\hat \sigma_{ee}(\tau,\omega)\rangle=\langle\hat \sigma_{e'e'}(\tau,\omega')\rangle\!=\!\langle\hat \sigma_{e'e'}(\tau,-\omega)\rangle, \\
\langle\hat \sigma_{eg}(\tau,\omega)\rangle=\langle\hat \sigma_{e'g'}(\tau,\omega')\rangle\!=\!\langle\hat \sigma_{e'g'}(\tau,-\omega)\rangle.
\end{split}
\end{equation}
Combining Eq.(\ref{sol_sym2}) and Eq.(\ref{sol_sym3}) we conclude that
\begin{equation}
\begin{split}
&\langle\hat \sigma_{ee}(\tau,\omega)\rangle=\langle\hat \sigma_{ee}(\tau,-\omega)\rangle, \\
&\langle\hat \sigma_{eg}(\tau,\omega)\rangle=-\langle\hat \sigma_{eg}(\tau,-\omega)\rangle.
\end{split}
\end{equation}
Therefore, if all the atoms are initially in the ground state, $\langle\hat \sigma_{ee}(t,\omega)\rangle$ is an even function of $\omega$ and $\langle\hat \sigma_{eg}(t,\omega)\rangle$ is odd.  All these symmetry properties can be checked against the results below.  We note that this statement is not true for arbitrary initial conditions or for the $\sigma$-operators themselves.

According to Eqs. (\ref{ee-DJ-second}) and (\ref{eg-DJ-second}), the differential equations for  $\sigma _{ee}^{i(n)} $  and  $\sigma _{eg}^{i(n)} $  are
\begin{eqnarray}
\frac{d \hat \sigma _{ee}^{i(n)}}{d \tau} &\!=\!& \frac{i}{2} \left[ {f(\tau )\hat \sigma _{eg}^{ i(n-1) } \!-\! f^{*}(\tau )\hat \sigma _{ge}^{ i(n-1) }  } \right]\! ,\label{ee(n)-DJ}\\
\frac{d \hat \sigma _{eg}^{i(n) }}{d\tau } &\!=\!& i\frac{{f^{*}(t)}}
{2}(2\hat \sigma _{ee}^{i(n - 1)} \! - \delta _{n1} ) + i\sum\limits_{j \ne i} k_{ij} \sum\limits_{p = 0}^{n - 1} {\hat \sigma _{eg}^{i(p)}\! \hat \sigma _{ee}^{j(n - p)} } \nonumber\\
&&\hspace{-0.mm} + i\sum\limits_{j \ne i} {k_{ij} } \hat \sigma _{eg}^{i(n)} \hat \sigma _{ee}^{j(0)} ,\label{eg(n)-DJ}
\end{eqnarray}
where $\delta _{n1}$ is the Kronecker delta function. 
To start the recurrence procedure, we need the initial operators $\hat\sigma_{ee}^{i(0)}$ and $\hat\sigma _{eg}^{i(0)}$. They are derived from Eqs. (\ref{ee-DJ-second}) and (\ref{eg-DJ-second}) for $\omega=0$
\begin{equation}\label{initial_sigma}
\begin{split}
&\hat\sigma_{ee}^{i(0)}(\tau)=\hat\sigma_{ee}^{i(0)}(\tau_0), \\
&\hat\sigma_{eg}^{i(0)}(\tau)=\hat\sigma_{eg}^{i(0)}(\tau_0)\hat Q^i(\tau-\tau_0),
\end{split}
\end{equation}
where $\tau_0$ is the initial time and $\hat Q^i$ is defined as
\begin{equation}\label{def_Q}
\hat Q^i(\tau)\equiv \exp \Bigg(i\;\tau \sum_{s \ne i}k_{is} \hat \sigma _{ee}^{s(0)}(\tau_0)\Bigg).
\end{equation}

The equivalent integral form of Eq. (\ref{ee(n)-DJ}) is 
\begin{align}\label{ee-IJ}
\hat \sigma _{ee}^{i(n)} (\tau) =&\frac{i}{2} \int \limits_{\tau _0 }^{\tau}  {d\tau _1 } \Big[f(\tau _1 ) \hat \sigma _{eg}^{i(n - 1)} (\tau _1 )\nonumber\\
&-f^{*}(\tau _1 ) \hat \sigma _{ge}^{i(n - 1)}(\tau _1 )\Big]  .
\end{align}
We get an equivalent integral form of Eq. (\ref{eg(n)-DJ}) in two steps. After multiplying
this equation  by $\hat Q^i(-\tau)$,
grouping terms with $\hat  \sigma _{eg}^{i(n)} $ and utilizing the time independence of $\hat \sigma _{ee}^{s(0)}$, another differential equation is obtained
\begin{align}\label{eg-DJfinal}
\frac{d}{d \tau}\left( {\hat \sigma _{eg}^{i(n)} \hat Q^i(-\tau)} \right) =&\Bigg[\frac{if^{*}(\tau _1 )}{2} 
(2\hat \sigma _{ee}^{i(n - 1)} - \delta _{n1} ) \nonumber \\
&\hspace{-8.mm}{}+ i\sum\limits_{j \ne i} k_{ij} \sum\limits_{p = 0}^{n - 1} \hat \sigma _{eg}^{i(p)} \hat \sigma _{ee}^{j(n - p)} \Bigg]\hat Q^i(\tau) .
\end{align}
The corresponding integral equation is
\begin{align}\label{eg-IJ}
\hat \sigma _{eg}^{i(n)} (\tau )&= \int\limits_{\tau _0 }^{\tau}  d\tau _1 \Bigg[ i\frac{f^{*}(\tau _1 )}
{2}(2\hat\sigma_{ee}^{i(n - 1)}(\tau _1 ) - \delta_{n1} ) \nonumber\\
&\hspace{-10.mm}{}+ i\sum_{j \ne i} k_{ij}\sum\limits_{p = 0}^{n - 1} \hat \sigma_{eg}^{i(p)}(\tau _1 )\hat \sigma_{ee}^{j(n - p)} (\tau _1 )\Bigg]\hat Q^i(\tau\!-\!\tau _1).
\end{align}
In our approach the recurrence relations  (\ref{ee-IJ}) and  (\ref{eg-IJ}) provide the way of calculating any order of the expansions. To calculate $\hat \sigma _{eg}^{i(n)}$, one needs to calculate $\hat \sigma _{ee}^{i(n)}$ first.  What we really want is the expectation values of the $\sigma$-operators, especially  $\left\langle {\hat \sigma _{ee} ^i} \right\rangle $. With the help of the recurrence relations, we can express all $\hat \sigma _{ee}^{i(n)} $ and  $\hat \sigma _{eg}^{i(n)} $ in terms of  $\hat \sigma _{ee}^{j(0)} $ and  $\hat \sigma _{eg}^{j(0)} $. Since we  can  easily find the expectation values of any product  of  $\hat \sigma _{ee}^{j(0)} $ and  $\hat \sigma _{eg}^{j(0)}$, in principle,  we can find exactly any  term in the  expansions (\ref{ee-expansion}) and (\ref{eg-expansion}).
In our calculation, we assume that all atoms are initially in the ground state. Consequently, the following  expectation values are necessarily equal to zero for any atom $i$
\begin{equation}\label{zeros}
\hspace{-0.mm}\left\langle\! {\hat \sigma _{eg}^{i(0)} }\!\!  \ldots \! \right\rangle \!=\!\left\langle\! \ldots {\hat \sigma _{ge}^{i(0)} }\right\rangle\!=\!\left\langle {\hat \sigma _{ee}^{i(0)} }\!\!  \ldots\!  \right\rangle
\!=\!\left\langle \! \ldots  {\hat \sigma _{ee}^{i(0)} }\right\rangle\!=\!0.
\end{equation}

The operator $\hat Q$ only produces phase factors. For example, for any $\hat a$, the following relation are satisfied
\begin{equation}\label{Qpropert}
\begin{split}
&\left\langle  \hat a \hat Q^{i}(\tau)  \right\rangle=\left\langle  \hat a \right\rangle,\\ 
&\left\langle  \hat a \hat Q^{i}(\tau) \hat \sigma_{eg}^{j(0)} \right\rangle=\exp(i\tau k_{ij})\left\langle  \hat a \hat \sigma_{eg}^{j(0)} \right\rangle.
\end{split}
\end{equation}
They follow from the definitions of $\hat Q$ and the $\sigma$-operators. Will use this property of $\hat Q$ to have all $k_{ij}$ in our formula exclusively in the phase factors $\exp(i\tau k_{ij})$. 

Since in the Heisenberg picture the wave function is time-independent, the average  $\left\langle \textnormal{ } \right\rangle $  can go through integrals contained in the recurrence relations. We find a new recurrence relation for $\left\langle {\hat \sigma _{ee}^{i(n)} } \right\rangle $ after substituting the expression (\ref{eg-IJ}) for $ \hat \sigma_{eg}^{i(n-1)}$  into Eq. (\ref{ee-IJ})
\begin{widetext}
\begin{equation}\label{ee(n)recurrence}
\left\langle \hat \sigma_{ee}^{i(n)} (\tau ) \right\rangle  =
{\mathrm{Re}}\left[
 \int\limits_{\tau_0}^{\tau}\!\! d{\tau_1} \frac{F(\tau) - F(\tau_1)}{2}\left(\!f^*(\tau_1)\left(\delta_{n2} - 2\left\langle \hat \sigma_{ee}^{i(n - 2)} (\tau_1 ) \right\rangle \right)
-\!\sum_{j \ne i} k_{ij} \sum\limits_{p = 0}^{n - 2}\! \left\langle\hat \sigma _{eg}^{i(p)} (\tau _1 )\hat \sigma _{ee}^{j(n - p)} (\tau _1 ) \right\rangle\!  \right)\! \right] ,
\end{equation}
\end{widetext}	
where  $F(\tau ) = \int_{\tau _0 }^\tau  {d\tau 'f(} \tau ')$.
For $n=2$, this equation, together with  expressions (\ref{zeros}), leads to   
\begin{equation}
\left\langle {\hat \sigma _{ee}^{i(2)} (\tau )} \right\rangle  = \int\limits_{\tau _0 }^{\tau}  {d\tau _1 } \int\limits_{\tau _1 }^\tau d\tau _2 \Phi(\tau_1 ,\tau_2) ,
\end{equation}
where
\begin{equation}
\Phi(\tau_1, \tau_2)=\frac{{f(\tau _1 )f^*(\tau _2 ) + f^*(\tau _1 )f(\tau _2 )}}{4}.
\end{equation}
For any symmetric function $\Psi(\tau _1 , \tau _2 )=\Psi( \tau _2 ,\tau _1 )$, the following relation is true (if the integrals exist)
\begin{equation*}
\begin{split}
\int\limits_{\tau _0 }^\tau  {d\tau _1 } \int\limits_{\tau _0 }^{\tau _1 } {d\tau _2 \Psi(\tau _1 ,\tau _2 )} &= \int\limits_{\tau _0 }^\tau  {d\tau _1 } \int\limits_{\tau _1 }^\tau  {d\tau _2 \Psi(\tau _1 ,\tau _2 )}\\
& = \frac{1}{2}\int\limits_{\tau _0 }^\tau  {d\tau _1 } \int\limits_{\tau _0 }^\tau  {d\tau _2 \Psi(\tau _1 ,\tau _2 )}\; .
\end{split}
\end{equation*}
Using this property, we finally get 
\begin{equation}\label{sigma_ee(2)}
\left\langle {\hat \sigma _{ee}^{i(2)} (\tau )} \right\rangle  = \frac{{\left| {F(\tau ) } \right|^2}} {4} \; .
\end{equation}
The last relation shows that the effects of interactions comes through higher orders than $n=2$. This, of course, would not be true if the initial conditions were different; in our case, since there are no excited atoms initially, there are no effects due to Rydberg-Rydberg interactions. For these initial conditions, the excitation always starts as isolated atom excitation. 

Besides Eq. (\ref{ee(n)recurrence}), we can derive more auxiliary relations for expectation values. They are useful if the same type of recurrence relations is repeatedly used. In our case, the expectation value $\left\langle \hat S(\tau') \hat \sigma _{ee}^{i(2)} (\tau )\right\rangle $, for an arbitrary operator $\hat S(\tau')$, will be particularly helpful. After substituting expression (\ref{ee-IJ}) for $n=2$, and then using Eq. (\ref{eg-IJ}) for $n=1$, we derive
\begin{align}\label{S_exp_val}
&\left\langle \hat S(\tau') \hat \sigma _{ee}^{i(2)} (\tau )\right\rangle = \frac{\left|F(\tau )\right|^2}{4}\left\langle \hat S(\tau') \right\rangle \nonumber\\
&\phantom{aaaaaaaaa}-\sum_{j\neq i}\frac{i k_{ij}}{4}\left\langle \hat S(\tau') \hat \sigma _{eg}^{i(0)} (\tau_0)\hat \sigma _{eg}^{j(0)} (\tau_0)\right\rangle\nonumber\\
&\phantom{aaaaaa}\times\int\limits_{\tau_0}^{\tau}d\tau_1 e^{i(\tau_1-\tau_0)k_{ij}} F(\tau_1)(F(\tau)-F(\tau_1))\; .
\end{align}
The complex conjugate of this formula leads to the expectation of $\left\langle \hat \sigma_{ee}^{i(2)}(\tau ) \hat S(\tau')\right\rangle$ for an arbitrary $\hat S(\tau')$ 
\begin{align}\label{conjS_exp_val}
&\left\langle\hat \sigma _{ee}^{i(2)}(\tau )\hat S(\tau')\right\rangle = \frac{\left|F(\tau )\right|^2}{4}\left\langle \hat S(\tau') \right\rangle \nonumber\\
&\phantom{aaaaa}+\sum_{j\neq i}\frac{i k_{ij}}{4}\left\langle \hat\sigma_{ge}^{j(0)}(\tau_0)\hat \sigma _{ge}^{j(0)} (\tau_0)\hat S(\tau') \right\rangle \nonumber\\
&\phantom{aaa}\times\int\limits_{\tau_0}^{\tau}\!d\tau_1 e^{-i(\tau_1-\tau_0)k_{ij}} F^*(\tau_1)(F^*(\tau)-F^*(\tau_1))\; .
\end{align}

Calculating $\left\langle {\hat \sigma _{ee}^{i(4)} (\tau )} \right\rangle$ takes more effort, so we split its contributions, given by Eq. (\ref{ee(n)recurrence}),  into three parts
\begin{equation}
\left\langle {\hat \sigma _{ee}^{i(4)} (\tau )} \right\rangle  =  - I_{41}  - I_{42}  - I_{43} \, . 
\end{equation}
The integrals are defined as follows
\begin{align}
&I_{41} =\!\int\limits_{\tau_0 }^{\tau}\! d{\tau_1}\frac{\left|F(\tau ) \right|^2}{4}
\mathrm{Re} \bigg[f^*(\tau _1 )  (F(\tau) - F(\tau_1))\bigg],\\
&I_{42} = \sum\limits_{j \ne i} k_{ij} \mathrm{Re} \Bigg[ \int\limits_{\tau _0 }^{\tau} d\tau_1\frac{F(\tau) \!-\! F(\tau_1)}{2}\nonumber\\
&\phantom{aaaaaaaaaaaaaaaa}\times\;  \left\langle \hat\sigma_{eg}^{i(1)} (\tau _1 )\hat \sigma_{ee}^{j(2)} (\tau _1 ) \right\rangle \Bigg] \; ,\\
&I_{43} = \sum\limits_{j \ne i} k_{ij} \mathrm{Re} \Bigg[\int\limits_{\tau _0 }^{\tau} d\tau_1 \frac{F(\tau) \!-\! F(\tau_1)}{2}\nonumber\\
&\phantom{aaaaaaaaaaaaaaaa}\times\;  \left\langle \hat \sigma_{eg}^{i(2)} (\tau _1 )\hat \sigma_{ee}^{j(1)} (\tau _1 )\right\rangle \Bigg] .
\end{align}
The integral $I_{41}$ can be calculated easily since  $\left\langle {\hat \sigma _{eg}^{i(2)} (\tau )} \right\rangle$ is given by Eq. (\ref{sigma_ee(2)}). After integration over $\tau_2$, we obtain
\begin{equation}
I_{41}=\frac{\left|F(\tau) \right|^4}{16}
-{\mathrm{Re}}\!\left[\frac{F(\tau)}{8}\int\limits_{\tau _0}^\tau\! {d\tau_1}\, g^*(\tau_1) F^2(\tau_1) \right].
\end{equation}

Only $I_{42}$ and $I_{43}$ include interactions. The calculation procedure is to apply Eqs. (\ref{ee-IJ}) and  (\ref{eg-IJ}) repeatedly, until the only operators left are  $\hat\sigma_{eg}^{i(0)}$ and $\hat\sigma_{ee}^{i(0)}$, whose expectation values are trivial to calculate. Using the auxiliary relations (\ref{ee(n)recurrence})-(\ref{S_exp_val}) often simplifies the derivation of expectation values. 
To evaluate $ \left\langle \hat\sigma_{eg}^{i(1)} (\tau _1 )\hat 
\sigma_{ee}^{j(2)} (\tau _1) \right\rangle $ in $I_{42}$, we can use Eq. (\ref{S_exp_val}). The result is 
\begin{equation}
I_{42}\! =\!\!\sum_{j \neq i}\! k_{ij}{\mathrm{Im}}\!\left[F(\tau )\!\!\int\limits_{\tau _0 }^{\tau}\!\!  d\tau_1 \frac{\left|F(\tau_1)\right|^2}{8}F^*(\tau _1)\! \right] .
 \end{equation}
This integral is canceled out by one of the $I_{43}$ terms, so there is no need to consider it in detail.

It is convenient to have $k_{ij}$ in the exponential (phase) factors only.  This can be done for the sum ${\mathcal{I}}_4=I_{42}+I_{43}$. After several partial integrations, the simplified form of ${\mathcal{I}}_4$ is
\begin{align}\label{final expansion}
{\mathcal{I}}_4 =&\frac{1}{4}
\sum\limits_{j \ne i} {\mathrm{Re}}\! \Bigg[\! \int\limits_{\tau_0 }^\tau\! d\tau_1 f(\tau _1)\left(F(\tau)\!-\!2F(\tau _1) \right) \nonumber\\
& \times\;\int\limits_{\tau_0}^{\tau_1}\!{d\tau _2 g^*(\tau _2)}F^*(\tau _2) \left( e^{i(\tau _1 - \tau _2 )k_{ij} }\! -\! 1 \right) \Bigg] .
\end{align}
For a system of $N$ atoms, one just needs to calculate these integrals for given $k_{ij}$. 

The derived expressions for $I_{41}$ and ${\mathcal{I}}_4$  are formally sufficient but they might not be very convenient in the limit of a large number of atoms $N$, where we replace the sum  $\sum\nolimits_{j \ne i} {} $ by the integral  $\int\rho {d^3 R} $, where $\rho$ is the atom density. For the terms we explicitely consider in our expansion,  it can be shown by direct calculation that this replacement is equivalent to averaging over all possible spatial distribution of atoms. For a large sample and an arbitrary pulse shape, an easier way is to first find the sum in Eq. (\ref{final expansion}) (i.e. the integral $\int\rho  {d^3 R} $). The last form of $I_{42}+I_{43}$ is very convenient to account for any angular dependence of $k_{ij}$.  As an example, we give the result for the excitation probabilities in large systems (surface effects are ignored) with uniform densities. For large homogeneous systems and  $k_{ij}=2 \pi C_s T /R^s$, where $C_s$ may be angular-dependent, we obtain
$$
\int {d^3 R}\;\left( e^{i(\tau _1 - \tau _2 )k_{ij} }\! -\! 1 \right) =\lambda (|C_s| T (\tau _1 - \tau _2 ))^{3/s}.
$$
In this expression, $\lambda$ is a parameter which depends on the type of interactions.
Note that $\tau _1 -\tau _2$ is always positive in ${\mathcal{I}}_4$ .

The ensemble averaged ${\mathcal{I}}_4$ for large homogeneous systems is 
\begin{align}\label{I4lagesamples}
{\mathcal{I}}_4 =&
\lambda \; \rho \;(C_s T)^{3/s}{\mathrm{Re}}\left[ \int\limits_{\tau_0 }^\tau\! d\tau_1 f(\tau _1)\left(F(\tau)-2F(\tau _1) \right)\right.\nonumber\\
&\left.\times\int\limits_{\tau_0}^{\tau_1}\!{d\tau _2 g^*(\tau_2)}F^*(\tau_2) (\tau_1 - \tau_2)^{3/s} \right].
\end{align}
For resonant excitation and real $f(t)$, the ensemble averaged expansion of excitation probabilities is 
\begin{equation}\label{prob-exp}
P_{\mathrm{exc}}  = \frac{\pi ^2}{4} \frac{I}{I_{\mathrm{sat}} } - \frac{\pi ^4}{48}
\left(1+ \gamma\rho \left| (C_s  \right|T )^{3/s} \right) \frac{I^2 }{I_{\mathrm{sat}}^2} + \ldots\, .
 \end{equation}
where $ I_{{\mathrm{sat}}}$ is the saturation laser intensity (for isolated atoms) and $T$ is the pulse duration. The introduction of $ I_{{\mathrm{sat}}}$ allows  using a single formula $P_{\mathrm{exc}}= \sin^2(\sqrt{ I/I_{\rm sat}} \pi/2 )$ for resonant excitation of isolated atoms regardless of the type of excitation pulses. For $s=6$ we have the van der Waals interactions and for $s=3$  the dipole-dipole interactions. The values of the parameter $\gamma$ in various cases of laser pulses and interactions are presented in Table \ref{gama_factor}.  
We note that the parameter $\gamma$ for the angular dependent dipole-dipole interactions (with aligned dipole moments) $V(R)=\frac{U_3}{R^3}\left(1-\cos^2\theta\right)$ is $\gamma=4\gamma'/3\sqrt{3}$, where $\gamma'$ corresponds to the isotropic interaction $U_3/R^3$. 

These formulae have the first contributions of interactions to excitation probabilities. Note that we never assumed that the interactions were weak. Actually, Eq. (\ref{final expansion}) is consistent with the limit $k_{ij}\rightarrow \infty$ for any $i,j$. In this limit, there is no contribution from the exponential terms since they oscillate infinitely fast. The remaining part reproduces exactly what one gets when the exact solution (\ref{strong_inter_solut}) in the limit $k_{ij}\rightarrow \infty$  is expanded in powers of $\omega$.
\begin{table}
\caption{\label{gama_factor}The parameter $\gamma$ in the expansion (\ref{prob-exp}) of excitation probabilities for various interaction potentials and excitation pulses.  The pulse envelope for a Gaussian pulse is $g(\tau)=e^{-\tau^2}$ and for a square pulse is $g(\tau)=\Theta(1-\tau)$.}
\begin{ruledtabular}
\begin{tabular}{|l|c|c|c|}
&\multicolumn{3}{c|}{$\gamma$ }\\ \cline{2-4}
 pulse type& $C_3/R^3$ & $\frac{U_3}{R^3}(1-3\cos^2 \theta)$ & $C_6/R^6$ \\
 \hline
Gauss. pulse&32.1138& 24.7212& 10.8627\\
 \hline
Square pulse&$\frac{2 \pi^3}{5}$&$\frac{8 \pi^3}{15\sqrt{3}}$&$\frac{128 \pi^2}{189}$\\
\end{tabular}
\end{ruledtabular}
\end{table}

\subsection{$\Omega$-Expansion of the correlation function} 
We also calculate the spatial correlation function $P(i,j)$ between different atoms $i$ and $j$
\begin{equation}\label{cor_fun}
P(i,j)=\frac{\left\langle \hat\sigma _{ee}^{i} (\tau )\hat\sigma _{ee}^{j} (\tau ) \right\rangle}{\left\langle \hat\sigma _{ee}^{i} (\tau )\right\rangle \left\langle \hat\sigma _{ee}^{i} (\tau )\right\rangle}.
\end{equation}
We first expand $c(i,j)\equiv\left\langle \hat\sigma _{ee}^{i} (\tau )\hat\sigma _{ee}^{j} (\tau ) \right\rangle$ in powers of $\omega$. Since $\left\langle \hat\sigma _{ee}^{i} (\tau )\hat\sigma _{ee}^{j} (\tau ) \right\rangle$ is essentially a probability, we expect only even terms in the expansion 
\begin{equation}\label{cor_fun_expans}
\left\langle \hat\sigma _{ee}^{i} (\tau )\hat\sigma _{ee}^{j} (\tau ) \right\rangle=\omega^2 c^{(2)}(i,j)+\omega^4 c^{(4)}(i,j)+\ldots\, ,
\end{equation}
where $c^{(0)}(i,j)=0$ due to the initial conditions (\ref{zeros}).
We can directly verify that the lowest odd $c^{(n)}(i,j)$ terms vanish.
According to Eq. (\ref{zeros}), $c^{(1)}(i,j)$ is equal to zero
\begin{equation}
c^{(1)}(i,j)=
\left\langle \hat\sigma _{ee}^{i(1)}\hat\sigma _{ee}^{j(0)}  \right\rangle+\left\langle \hat\sigma _{ee}^{i(0)} \hat\sigma _{ee}^{j(1)}  \right\rangle=0.
\end{equation}
Using  Eq. (\ref{ee-IJ}), we also find
\begin{equation}\label{c2term}
c^{(2)}(i,j)=\left\langle \hat\sigma _{ee}^{i(1)} (\tau )\hat\sigma _{ee}^{j(1)} (\tau ) \right\rangle=\frac{\left|F(\tau) \right|^2}{4} \delta_{ij}=0 ,
\end{equation}
since  $i\neq j$. Relation (\ref{sigma_ee(2)}) also follows from the last formula because $\hat\sigma_{ee}^{i}=\hat\sigma_{ee}^{i}\hat\sigma_{ee}^{i}$. We can explicitly show that $c^{(3)}(i,j)=\left\langle \hat\sigma _{ee}^{i(1)} (\tau )\hat\sigma _{ee}^{j(2)} (\tau ) \right\rangle+\left\langle \hat\sigma _{ee}^{i(2)} (\tau )\hat\sigma _{ee}^{j(1)} (\tau ) \right\rangle=0$ using Eqs. (\ref{S_exp_val}) and (\ref{conjS_exp_val}) with $\hat S=\hat\sigma _{ee}^{i,j(1)}$ and $\left\langle\hat\sigma _{ee}^{i,j(1)}\right\rangle=0$.  Therefore,  $c^{(4)}(i,j)$ is the first nontrivial term in the expansion of $\left\langle \hat\sigma _{ee}^{i} (\tau )\hat\sigma _{ee}^{j} (\tau ) \right\rangle$
\begin{align}\label{c4term}
c^{(4)}(i,j)=&\left\langle \hat\sigma _{ee}^{i(1)} (\tau )\hat\sigma _{ee}^{j(3)} (\tau ) \right\rangle+\left\langle \hat\sigma _{ee}^{i(2)} (\tau )\hat\sigma _{ee}^{j(2)} (\tau ) \right\rangle\\
&{}+\left\langle \hat\sigma _{ee}^{i(3)} (\tau )\hat\sigma _{ee}^{j(1)} (\tau ) \right\rangle.\nonumber
\end{align}
This is expected since the initial excitation probability of any atom is $\sim \omega^2$.
The first and last term in the right-hand side of the last equation are very similar. If we know one of them, it is easy to find the other one.
To find $\left\langle \hat\sigma _{ee}^{i(2)} (\tau )\hat\sigma _{ee}^{j(2)} (\tau ) \right\rangle$, we apply Eq. (\ref{S_exp_val}) for $\hat S(\tau)=\hat\sigma_{ee}^{i(2)}(\tau)$
\begin{align}
&\left\langle \hat \sigma_{ee}^{i(2)}(\tau) \hat \sigma _{ee}^{i(2)} (\tau )\right\rangle = \frac{\left|F(\tau )\right|^4}{16} \nonumber\\
&\phantom{aaaaaaa}-\sum_{s\neq j}\frac{i k_{js}}{4}\left\langle \hat \sigma_{ee}^{i(2)}(\tau) \hat \sigma _{eg}^{j(0)} (\tau_0)\hat \sigma _{eg}^{s(0)} (\tau_0)\right\rangle \nonumber\\
&\phantom{aaaaaaa}\times\int\limits_{\tau_0}^{\tau}d\tau_1 e^{i(\tau_1-\tau_0)k_{js}} F(\tau_1)(F(\tau)-F(\tau_1))\; .
\end{align}
The expectation value in the last formula is easily evaluated using Eq. (\ref{conjS_exp_val})
together with  the following identity
\begin{equation*}
\begin{split}
\left\langle \hat\sigma_{ge}^{q(0)}(\tau_0)\hat \sigma _{ge}^{i(0)}(\tau_0)\hat \sigma _{eg}^{j(0)} (\tau_0)\hat \sigma _{eg}^{s(0)} (\tau_0) \right\rangle=\delta_{is}\delta_{jq},
\end{split}
\end{equation*}
where we assume $i\neq j$.
The result is therefore
\begin{align}\label{c4term22}
\left\langle \hat \sigma_{ee}^{i(2)}(\tau) \hat \sigma _{ee}^{j(2)} (\tau )\right\rangle &= \frac{\left|F(\tau )\right|^4}{16}\nonumber\\ 
&\hspace{-20 mm}+\frac{k_{ij}^2}{16}\!\left|\int\limits_{\tau_0}^{\tau}\!d\tau_1 e^{i\tau_1k_{ij}} F(\tau_1)(F(\tau)\!-\!F(\tau_1))\right|^2\; .
\end{align}
A more convenient form obtained after partial integration is
\begin{align}\label{c4term22new}
\left\langle \hat \sigma_{ee}^{i(2)}(\tau) \hat \sigma _{ee}^{j(2)} (\tau )\right\rangle &= \frac{\left|F(\tau )\right|^4}{16}\nonumber\\ 
&\hspace{-20 mm}+\frac{1}{16}\!\left|\int\limits_{\tau_0}^{\tau}\!d\tau_1 e^{i\tau_1k_{ij}} F(\tau_1)(F(\tau)\!-2\!F(\tau_1))\right|^2. 
\end{align}

Instead of using Eqs. (\ref{ee-IJ})-(\ref{eg-IJ}), another approach to find $\left\langle \hat\sigma _{ee}^{i(1)} (\tau )\hat\sigma _{ee}^{j(3)} (\tau ) \right\rangle $ is to utilize the identity $\hat\sigma_{ee}^{j}=\hat\sigma_{ee}^{j}\hat\sigma_{ee}^{j}$, which gives
$\hat\sigma_{ee}^{j(3)}=\hat\sigma_{ee}^{j(0)}\hat\sigma_{ee}^{j(3)}+\hat\sigma_{ee}^{j(1)}\hat\sigma_{ee}^{j(2)}+\hat\sigma_{ee}^{j(2)}\hat\sigma_{ee}^{j(1)}+\hat\sigma_{ee}^{j(3)}\hat\sigma_{ee}^{j(0)}$. The expectation values of the first and the last term in this expansion vanish so that 
\begin{align}\label{c4term13}
\left\langle \hat\sigma _{ee}^{i(1)} (\tau )\hat\sigma _{ee}^{j(3)} (\tau ) \right\rangle
&=\left\langle \hat\sigma _{ee}^{i(1)} (\tau )\hat\sigma _{ee}^{j(1)}(\tau )\hat\sigma _{ee}^{j(2)} (\tau ) \right\rangle\nonumber\\
&{}+\left\langle \hat\sigma _{ee}^{i(1)} (\tau )\hat\sigma _{ee}^{j(2)}(\tau )\hat\sigma _{ee}^{j(1)} (\tau ) \right\rangle.
\end{align}
To evaluate $\left\langle \hat\sigma _{ee}^{i(1)} (\tau )\hat\sigma _{ee}^{j(1)}(\tau )\hat\sigma _{ee}^{j(2)} (\tau ) \right\rangle$, we first apply the recurrence relation (\ref{ee-IJ}) on $\sigma _{ee}^{i(1)} (\tau )$ and $\sigma _{ee}^{j(1)} (\tau )$
\begin{align}
\left\langle \hat\sigma _{ee}^{i(1)} (\tau )\hat\sigma _{ee}^{j(1)}(\tau )\hat\sigma _{ee}^{j(2)} (\tau ) \right\rangle\!&=\!\left\langle \hat\sigma_{ge}^{i(0)} (\tau_0 )\hat\sigma_{ge}^{j(0)}(\tau_1)\hat\sigma _{ee}^{j(2)} (\tau ) \right\rangle\nonumber\\
&\hspace{-19mm}\times\frac{-F^*(\tau)}{4}\! \int\limits_{\tau_0}^{\tau}d\tau_1 e^{-i(\tau_1-\tau_0)k_{ij}} f^*(\tau_1).
\end{align}
To get the expectation value in the right-hand side of the last equation, we use Eq. (\ref{S_exp_val}) together with the following  identity 
\begin{equation*}
\begin{split}
\left\langle \hat\sigma_{ge}^{i(0)}(\tau_0)\hat \sigma _{ge}^{j(0)}(\tau_0)\hat \sigma _{eg}^{j(0)} (\tau_0)\hat \sigma _{eg}^{s(0)} (\tau_0) \right\rangle=\delta_{is}.
\end{split}
\end{equation*}
The result is 
\begin{align}\label{c4term13solv}
\left\langle \hat\sigma_{ee}^{i(1)}(\tau)\hat\sigma_{ee}^{j(1)}(\tau ) \hat\sigma_{ee}^{j(2)} (\tau )\right\rangle &=\frac{i\;k_{ij}}{16}F^*(\tau )\!\int\limits_{\tau_0}^{\tau} \!d\tau_1 f^*(\tau_1)\nonumber\\ 
&\hspace{-23mm}\!\int\limits_{\tau_0}^{\tau}\!d\tau_2\;e^{i(\tau_2-\tau_1)k_{ij}}F(\tau_2)(F(\tau)-F(\tau_2)).
\end{align}
This can be additionally transformed using partial integration to get
\begin{align}\label{c4term13final}
\left\langle \hat\sigma_{ee}^{i(1)}(\tau)\hat\sigma_{ee}^{j(1)}(\tau ) \hat\sigma_{ee}^{j(2)} (\tau )\right\rangle &=\frac{-F^*(\tau )}{16}\!\int\limits_{\tau_0}^{\tau}\!\!d\tau_1 f^*(\tau_1)\nonumber\\
&\hspace{-25mm}\!\int\limits_{\tau_0}^{\tau}\!\!d\tau_2e^{i(\tau_2-\tau_1)k_{ij}} F(\tau_2)(F(\tau)-2F(\tau_2)). 
\end{align}
The only remaining term to calculate is the last term in Eq. (\ref{c4term13}).  From Eq. (\ref{ee-IJ}), we immediately find 
\begin{align}\label{c4term13new}
\left\langle \hat\sigma _{ee}^{i(1)} (\tau )\hat\sigma _{ee}^{j(2)}(\tau )\hat\sigma _{ee}^{j(1)} (\tau ) \right\rangle&=\nonumber\\
&\hspace{-25mm}\frac{\left|F(\tau)\right|^2}{4}\left\langle \hat\sigma _{ge}^{i(0)} (\tau_0)\hat\sigma _{ee}^{j(2)}(\tau)\hat\sigma _{eg}^{j(0)} (\tau ) \right\rangle.
\end{align}
We proceed using Eq. (\ref{ee-IJ}) to express $\hat\sigma _{ee}^{j(2)}(\tau )$ in terms of $\hat\sigma _{eg}^{j(1)}$  and $\hat\sigma _{ge}^{j(1)}$,  and then apply Eq. (\ref{eg-IJ}) on $\hat\sigma _{eg}^{j(1)}$  and $\hat\sigma _{ge}^{j(1)}$.  The simplified form after these substitutions is 
\begin{align}\label{c4term13newsolv}
&\left\langle \hat\sigma _{ee}^{i(1)} (\tau )\hat\sigma _{ee}^{j(2)}(\tau )\hat\sigma _{ee}^{j(1)} (\tau ) \right\rangle=\frac{-i\left|F(\tau)\right|^2}{16}\nonumber\\
&\phantom{aa}\times k_{ij}\!\int\limits_{\tau_0}^{\tau}\!\!d\tau_1 f^*(\tau_1)\int\limits_{\tau_0}^{\tau_1}d\tau_2 e^{i(\tau_2-\tau_1)k_{ij}}F(\tau_2).
\end{align}
Partial integration will leave $k_{ij}$ only in the phase factor
\begin{align}\label{c4term13newfinal}
&\left\langle \hat\sigma _{ee}^{i(1)} (\tau )\hat\sigma _{ee}^{j(2)}(\tau )\hat\sigma _{ee}^{j(1)} (\tau ) \right\rangle\!=\!\frac{-\left|F(\tau)\right|^2}{16}\!\int\limits_{\tau_0}^{\tau}\!\!d\tau_1 f^*(\tau_1)F(\tau_1)\nonumber\\
&\phantom{aaaa} +\!\frac{\left|F(\tau)\right|^2}{16}\! \!\int\limits_{\tau_0}^{\tau}\!\!d\tau_1 f^*(\tau_1)\int\limits_{\tau_0}^{\tau_1}\!\!d\tau_2 e^{i(\tau_2-\tau_1)k_{ij}}F(\tau_2).
\end{align}
Combining Eqs.  (\ref{c4term22new}), (\ref{c4term13final})  and (\ref{c4term13newfinal}), we get our final formula
\begin{equation}\label{c4termfinal}
c^{(4)}(i,j)=\frac14\left|\int\limits_{\tau_0}^{\tau}\!\!d\tau_1 e^{i\tau_1 k_{ij}}f(\tau_1)F(\tau_1)\right|^2.\\
\end{equation}
The first term in the expansion of the product $\left\langle \hat\sigma _{ee}^{i} (\tau )\right\rangle \left\langle \hat\sigma _{ee}^{i} (\tau )\right\rangle$ is also proportional to $\omega^4$. According to Eq. (\ref{sigma_ee(2)}), it is
\begin{equation}\label{c2product}
\left\langle \hat\sigma _{ee}^{i} (\tau )\right\rangle \left\langle \hat\sigma _{ee}^{i} (\tau )\right\rangle=\frac{\left|F(\tau)\right|^4}{16}\omega^4+\ldots\;.\\
\end{equation}
From Eqs. (\ref{cor_fun}), (\ref{c4termfinal}) and (\ref{c2product}), we find the correlation function 
for low laser power to be
\begin{equation}\label{cor_fun_final}
P(i,j)=\frac{4\left|\int\limits_{\tau_0}^{\tau}\!\!d\tau_1 e^{i\tau_1 k_{ij}}f(\tau_1)F(\tau_1)\right|^2}{\left|F(\tau)\right|^4}.
\end{equation}
This simple formula includes the effects of interactions, frequency detuning, and possible frequency chirp. As pointed out in \cite{Hernandez,Stanojevic07}, the numerical calculation of the correlation function is much more demanding than the calculation of the excitation probability in the superatom approach. Both calculations include running simulations for many random spatial distributions of atoms to find ensemble averaged probabilities and correlation functions.  In the superatom approach, the conceptual issue in the calculation of the  correlation function is related to the fact that superatoms are extended objects. The question is how much the numerical correlation function averaged over many random spatial distributions of atoms is accurate over distances less that the average size of a superatom.

\begin{figure}[b]
    \centerline{\epsfxsize=3.2 in\epsfclipon\epsfbox{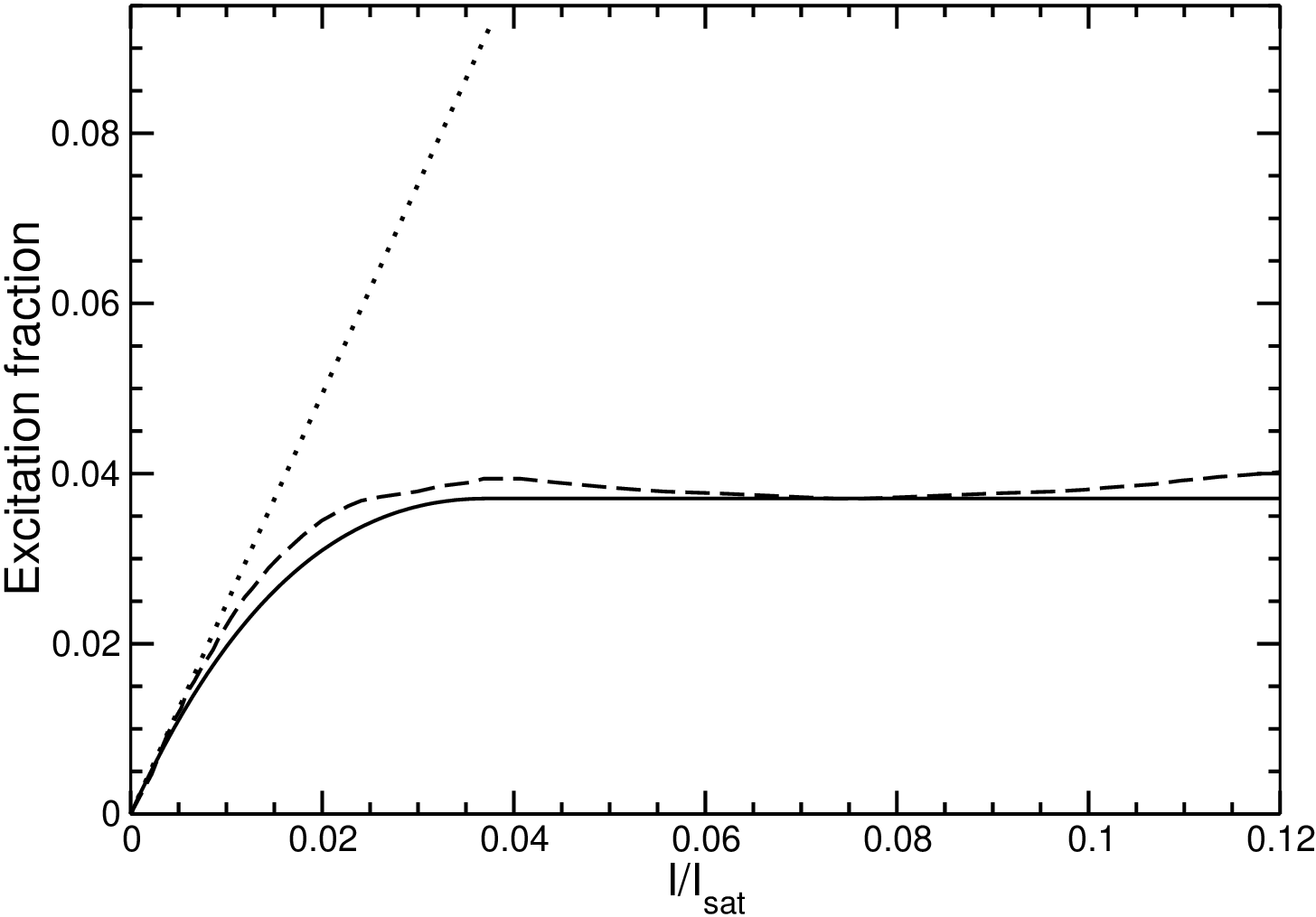}}
\caption{\label{satur_exc} Rydberg excitation fraction as a function
of laser intensity $I$. The solid line is obtained using  Eq. (\ref{P_0new}) for  $I> I_0$ and  the model dependence (\ref{model_depend}) for $I<I_0$. The result is comparable with the numerical solution  (dashed line) of the many-body wave function \cite{Hernandez}. The dotted line is the noninteracting limit. The parameters used for this simulation are $C_6= 2.64 \times 10^{22}\times 7/60$, $\rho=6.5$ $10^{10}$ cm$^{-3}$, and a laser bandwidth $\Gamma=120$ MHz. }
\end{figure}

\section{Results and Discussion}
We know from theoretical \cite{tong04,Hernandez} and experimental studies \cite{tong04,singer04,Liebisch} that excitation fractions are easily saturated in large systems of strongly interacting atoms. This, of course, cannot be concluded from the first two terms in Eq. (\ref{prob-exp}). However, assuming that the probabilities are saturated, we can use the maximum excitation probability, determined from these two terms to estimate the saturated excitation fraction. From the corresponding laser irradiance, we can get an estimate of the saturation intensity $I_0$ for which the excitation fraction becomes saturated due to interactions.
\begin{figure}[b]
    \centerline{\epsfxsize=3.2 in\epsfclipon\epsfbox{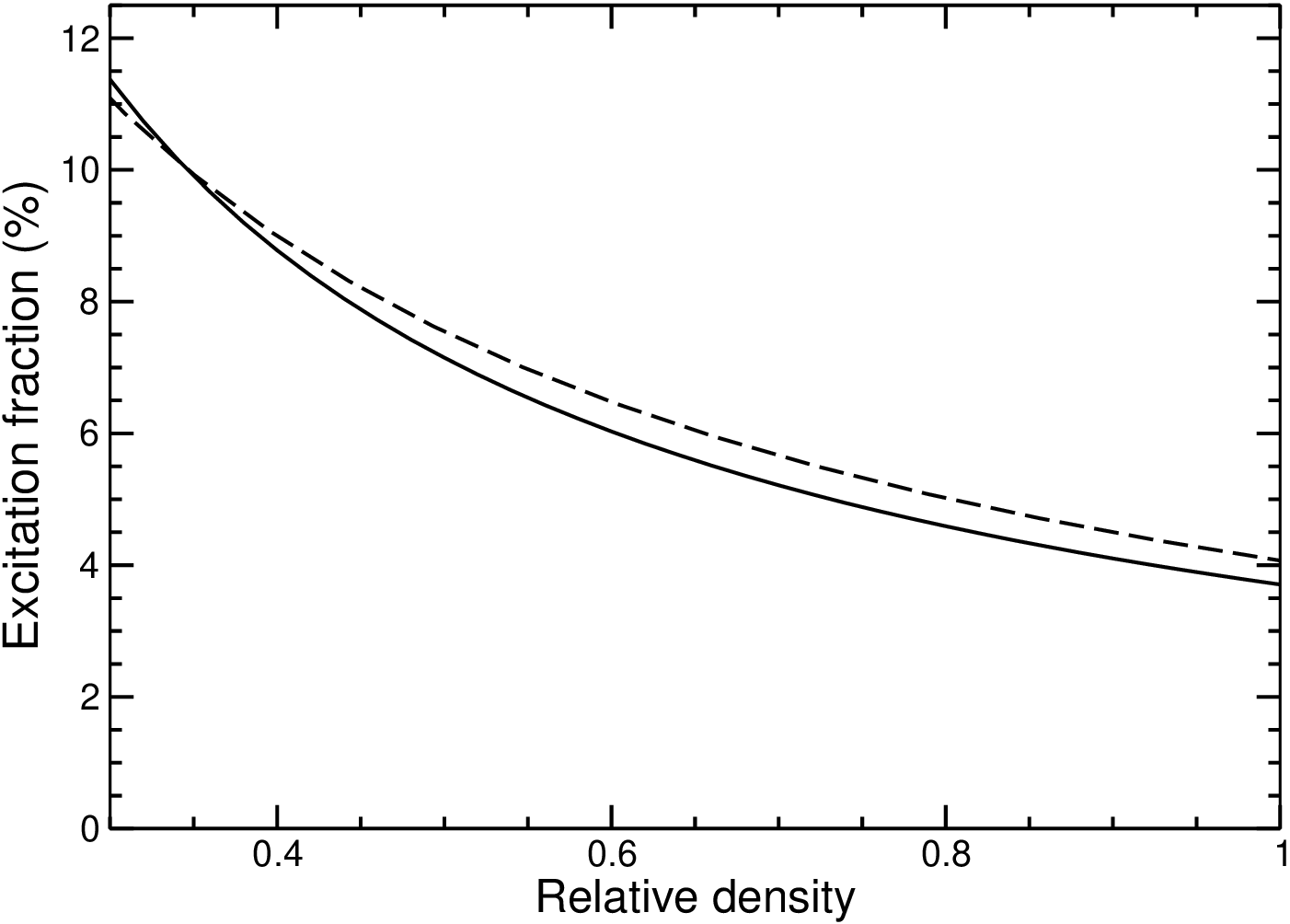}}
\caption{\label{dens_dep} Density dependence of the Rydberg excitation fraction. The solid line is this dependence obtained using Eq. (\ref{P_0new}) and the dashed line is  obtained using  a mean field model  described in \cite{tong04} for $I/I_{\rm sat}=0.2$.  The values of $C_6$ and $\Gamma$ in this figure and  Fig. \ref{satur_exc} are the same (the pulse duration is adjusted to get this bandwidth). The highest atom density $\rho_{\rm max}$ in this figure corresponds to the atom density in Fig. \ref{satur_exc}.}
\end{figure}
\begin{figure}[t]
    \centerline{\epsfxsize=3.2 in\epsfclipon\epsfbox{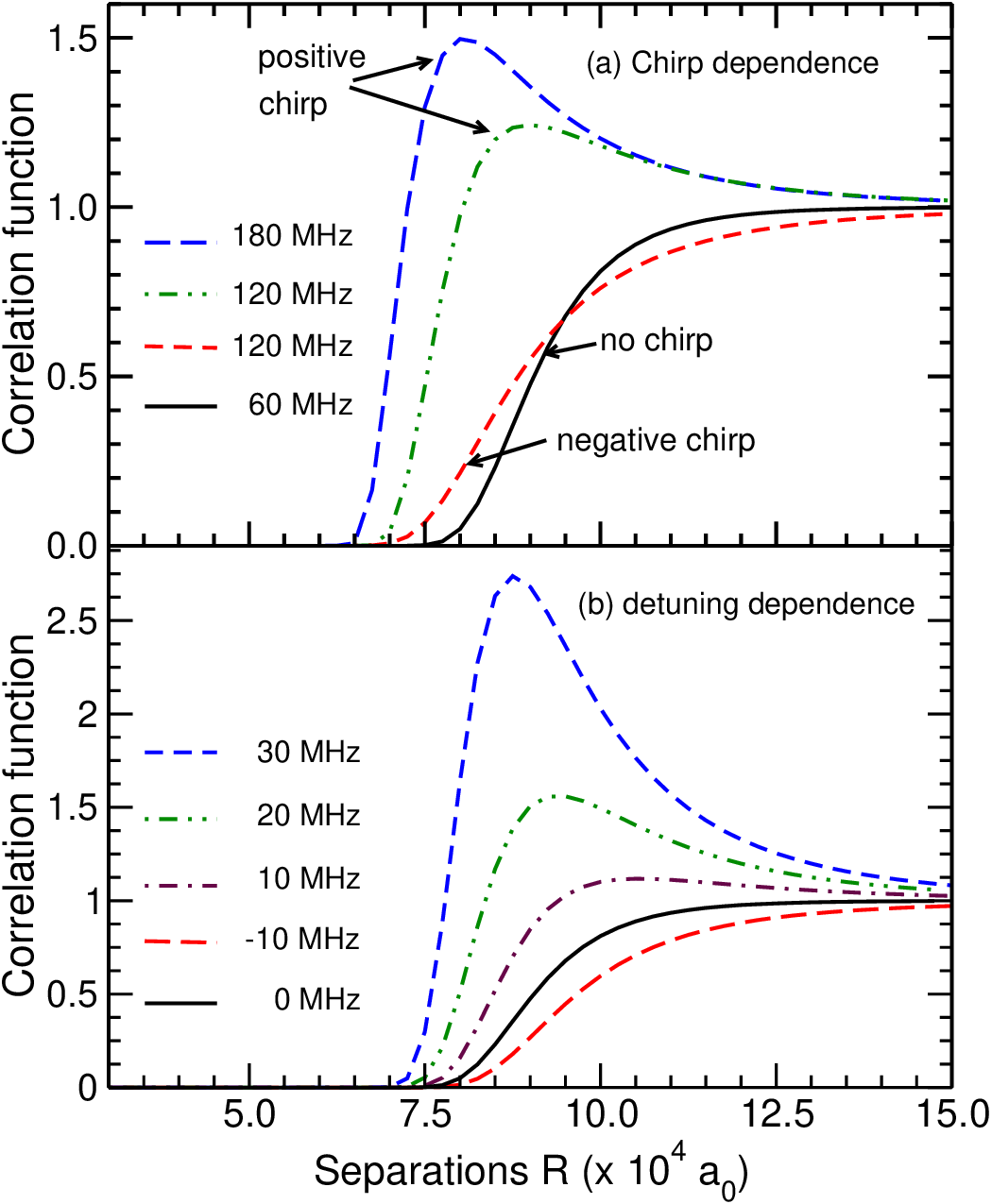}}
\caption{\label{corfun}Correlation function between excited atoms for various detunings and chirp parameters.  In (a), the laser bandwidth $\Gamma$  was varied as indicated, assuming that the excess bandwidth is caused by linear chirp. The solid line ($\Gamma=60$ MHz) represents the dependence with no chirp. The dashed line crossing the solid one is the only  dependence with a negative chirp. Characteristic particle correlations exist only for a positive chirp. In (b), the frequency detuning   was varied as indicated, assuming no chirp ($\Gamma=60$ MHz). Since the interactions are attractive, characteristic particle correlations exist only for positive detunings. This figure demonstates the strong dependence of the correlation function on detuning. }
\end{figure}

We introduce a supression factor $N_d$ defined as the ratio of the maximum excitation probabilities for isolated atoms and for interacting atoms. This $N_d$ is basically the number of atoms in the largest possible region containing no more than  one excited atom. The maximum excitation probability $P_0$ obtained directly from the first two terms in Eq. (\ref{prob-exp}) is 
\begin{equation}\label{P_0}
P_0\approx\frac{3/4}{1+ \gamma\rho (|C_s|T)^{3/s}}.
\end{equation}
Obviously, this formula underestimates the excitation probability in the absence of interactions $C_s\rightarrow0$ since  it gives $P_0=3/4$ instead of $P_0=1$. This is due to the truncation of the expansion. Also, in the limit $C_s\rightarrow\infty$, the first two terms in the expansion of $P_{\mathrm {exc}}$, according to  Eq. (\ref{strong_inter_solut}), lead to the estimate $P_0=3/4N$ instead of $P_0=1/N$. It seems that, in the first approximation, the limitations due to using truncated expansions could be overcome by replacing the factor $3/4$ by 1. This replacement leaves the estimated excitation fraction, as a function of the interaction parameters, basically the same. Another procedure, based on  Eq. (\ref{strong_inter_solut}), leads to the same correction.
We can use a modification of  Eq. (\ref{strong_inter_solut}) to get a model dependence which is correct in both limits of $C_s$. The simplest modification is
\begin{equation}\label{model_depend}
P_{\mathrm {exc}}\approx\sin^2\left(\sqrt{N_d I/I_{sat}} \pi/2 \right)/N_d .
\end{equation}
This $N_d$ is equal to one for isolated atoms and $N_d=N$ for a fully excitation blockade. For arbitrary interactions, $N_d$ is obtained requiring that the first two terms in the expansion of $P_{\mathrm {exc}}$ given by Eq. (\ref{model_depend}) match the evaluated terms in Eq. (\ref{prob-exp}). Strictly speaking, this model dependence can be only used for low laser power, or more precisely, up to the first maximum of the excitation probability. Due to the saturation nature of the excitation probability, we assume that this maximum is a reasonable estimate of $P_0$, even for somewhat higher laser power. The excitation fraction obtained using these assumptions agrees well with the numerical solution \cite{Hernandez} of the many-body wave function, as shown  in Fig. \ref{satur_exc}. Therefore, the estimates of the saturated fraction  $P_0$ and the saturation intensity  $I_0$ for interacting atoms are 
\begin{equation}\label{P_0new}
P_0=\frac{I_0}{I_{sat}}=\frac{1}{N_d}=\frac{1}{1+\gamma\rho (|C_s||T)^{3/s}}.
\end{equation}
This formula can be also viewed as a density dependence of the saturated excitation fraction. This dependence is shown  in Fig. \ref{dens_dep} for $C_6= 2.64 \times 10^{22}\times 7/60$ and a laser bandwidth $\Gamma=120$ MHz. The largest atom density $\rho_{\rm max}$ in Fig. \ref{dens_dep} is $6.5\times10^{10}$ cm$^{-3}$. In this figure, we also show the result from a mean-field model described in \cite{tong04}. The laser power used in the mean-field calculation was $I/I_{\rm sat}=0.2$. The agreement is fairly good. For the the above values of $C_6$ and $\Gamma$, and $\rho=\rho_{\rm max}$,  we also have the result on saturated excitation fraction from many-body simulations \cite{Hernandez}.  The saturated excitation fraction from the full numerical calculations slowly varies from 3.7 to 4.0 \%, while  we get $P_0=3.7$ \% using Eq. (\ref{P_0new}). The same parameters were used to get the Rydberg excitation fraction in Fig. \ref{satur_exc}. The only difference is that we adjust the pulse width to get the experimental $\Gamma=120$ MHz, while in \cite{tong04} and \cite{Hernandez} this bandwidth was caused by a linear chirp. 
We can compare the prediction of our formula (\ref{P_0new}) with the results from the numerical treatment \cite{Hernandez} for the experimental parameters in \cite{singer04}. The simulation \cite{FRunpublished} was done for $C_6= 4.97 \times 10^{22}$,  $T=37.5$ ns and a sample density of $2\times10^{9}$ cm$^{-3}$. The resonant excitation fraction was about 7.5 \%. For the same parameters, we get 8.2 \% using Eq. (\ref{P_0new}). It is interesting that Eq. (\ref{P_0new}) gives results  very similar to the results of the full numerical treatment in both simulations, even though the densities and pulse durations are very different.

In Fig. \ref{corfun}(a) we show the correlation function obtained from Eq. (\ref{cor_fun_final}) for various laser bandwidths $\Gamma$ caused by a linear frequency chirp. The  pulse width of a Gaussian pulse is $7.3$ ns and the interaction is characterized by $C_6= 2.64 \times 10^{22}\times 7/60$, the same parameters used in \cite{tong04} and the simulation \cite{Hernandez}. The dependence for $\Gamma=120$ MHz can be compared with the numerically obtained dependence in \cite{Hernandez} for the lowest optical field amplitude. These two dependences are remarkably similar. This figure confirms that there is a region where the correlation function is greater than one only for a positive chirp (the laser frequency linearly increases during excitation). In Fig. \ref{corfun}(b) we show how the correlation function is affected by frequency detuning. This figure demonstrates that the correlation function is very sensitive to frequency detuning. Although the  detunings  used in this figure are just a fraction of the laser bandwidth, the correlation function dependence varies significantly. Positive detunings partially compensate the effect of attractive interactions so that certain separations between excited pairs are preferable for a given positive detuning. Therefore, the correlation function is greater than one at such separations.  This is not the case for negative detunings. These properties are also shown in  numerical simulations \cite{Hernandez}. From these figures one may expect that relative small negative detunings can easily cancel out the effect of a possible positive chirp, which is easy to show using Eq. (\ref{cor_fun_final}). It is worth noting that the relatively simple formula (\ref{cor_fun_final}) reproduces well the main results of  demanding numerical calculations.

Even though our formula for excitation probabilities (\ref{P_0new}) and the correlation function (\ref{cor_fun_final}) are primarily derived for lower laser power, they describe related physical phenomena reasonably well despite their simplicity. Coupled with the fact that numerical many-body calculations are very demanding, these relatively simple formulae can be useful for actual experiments.

\section{\label{sec:conclusionEXP}Conclusion}
We have investigated the many-body excitation dynamics in ultracold Rydberg systems using the $\Omega$-expansion.
We have shown that the equations of motion can be solved by expanding the $\sigma$-operators in powers of  $\Omega$. Different terms in the power expansions of the $\sigma$-operators can be calculated using the recurrence relations. These recurrence relations are used to evaluate the expansion of the excitation probability  up to the $\Omega^4$ term.
For homogeneous large samples, the expansions  obtained are additionally ensemble averaged  in various cases of excitation pulses and interactions. We have also derived an explicit form of the correlation function between excited atoms for laser power and compared it with the recent numerical calculations. Our derived formulae agree well with recent numerical many-body simulation \cite{Hernandez}. Their simplicity makes them easily applicable in a range of experimental conditions 

\begin{acknowledgments}
We thank F. Robicheaux for fruitful discussions. This research was funded by the National Science Foundation.
\end{acknowledgments}


\end{document}